%% file: main.tex
\documentclass{article}
\usepackage{spconf,amsmath,graphicx}

\usepackage{xcolor}
\usepackage[printonlyused,nolist]{acronym}
\usepackage{amssymb}
\usepackage{booktabs}
\usepackage{gensymb}
\usepackage{multirow}
\usepackage{tikz}
\usepackage{pgfplots}
\usepackage{environ}
\usepackage{tikzscale}
\usepackage{subcaption}
\usepackage[skip=5pt]{caption}
\usepackage{blindtext}
\usepackage{setspace}
\usepackage{array}
\usepackage{cite}
\usepackage{amsfonts} 
\DeclareMathSymbol{\shortminus}{\mathbin}{AMSa}{"39}

\newcommand{\crunet}{$\mathbb{C}$RUNet}
\newcommand{\D}{D}

\begin{acronym}
    \acro{2D}{two-dimensional}
    \acro{3D}{three-dimensional}
    \acro{ANN}{artificial neural network}
    \acro{COSPA}{Complex-valued Spatial Autoencoder}
	\acro{DNN}{Deep Neural Network}
    \acro{DoA}{direction of arrival}
    \acro{DSB}{delay-and-sum beamformer}
    \acro{EaBNet}{Embedding and Beamforming Network}
    \acro{ESTOI}{extended Short-Time Objective Intelligibility}
	\acro{GRU}{gated recurrent unit}
    \acro{iCOSPA}{informed COSPA}
    \acro{MC}{multichannel}
    \acro{MVDR BF}{minimum variance distortionless response-beamformer}
	\acro{PESQ}{Perceptual Evaluation of Speech Quality}
	\acro{ReLU}{rectified linear unit}
	\acro{SDR}{signal-to-distortion ratio}
	\acro{SIR}{signal-to-interference ratio}
	\acro{SNR}{signal-to-noise ratio}
    \acro{STOI}{Short-Time Objective Intelligibility}
	\acro{STFT}{short-time Fourier transform}
    \acro{TSE}{target speaker extraction}
\end{acronym}

\title{Exploiting spatial information with the informed complex-valued spatial autoencoder for target speaker extraction}

\name{Annika Briegleb \qquad Mhd Modar Halimeh$^{\star}$\thanks{$^{\star}$M. M. Halimeh is now with Fraunhofer Institute for Integrated Circuits~IIS, Am Wolfsmantel 33, 91058 Erlangen, Germany.} \qquad Walter Kellermann\thanks{This work has been accepted to IEEE ICASSP 2023.}}
\address{Friedrich-Alexander-Universit\"at Erlangen-N\"urnberg, Erlangen, Germany\\
\{annika.briegleb, mhd.m.halimeh, walter.kellermann\}@fau.de}

\begin{document}

\ninept
\maketitle
\begin{abstract}
In conventional multichannel audio signal enhancement, spatial and spectral filtering are often performed sequentially. In contrast, it has been shown that for neural spatial filtering a joint approach of spectro-spatial filtering is more beneficial. In this contribution, we investigate the spatial filtering performed by such a time-varying spectro-spatial filter.
We extend the recently proposed complex-valued spatial autoencoder (COSPA) for the task of target speaker extraction by leveraging its interpretable structure and purposefully informing the network of the target speaker's position.
We show that the resulting informed COSPA (iCOSPA) effectively and flexibly extracts a target speaker from a mixture of speakers. We also find that the proposed architecture is well capable of learning pronounced spatial selectivity patterns and show that the results depend significantly on the training target and the reference signal when computing various evaluation metrics.
\end{abstract}
\section{Introduction}
\label{sec:intro}
While neural networks represent the state of the art for single-channel audio signal enhancement for some time now, they are only recently moving into the focus for multichannel audio signal enhancement and, hence, spatial filtering. There have been several approaches to guide spatial filters, i.e., beamformers, by estimating intermediate quantities by neural networks \cite{Zhang2017, Wang2018, Martin2020, Masuyama2020, Wang2020}. Other approaches construct a beamformer by estimating its weights by a neural network \cite{Halimeh2022, Tesch2022_2, Li2022, Meng2017, Xiao2016, Li2016} or replace the beamforming process by a neural network that directly estimates the clean speech signal \cite{Tan2022, Wang2020_2}. The first approach stays with the conventional definitions of beamformers, whereas the second and third approach exploit the nonlinear processing performed by the neural network and are denoted as neural spatial filters. 

In this paper, we focus on those neural spectro-spatial filters that estimate the beamformer weights by a neural network. Such neural spectro-spatial filters learn a spatially selective pattern for signal denoising in scenarios where only one speech source is active \cite{Halimeh2022, Tesch2022}. In this contribution, we extend one of such neural spectro-spatial filters, the \ac{COSPA} \cite{Halimeh2022}, for the problem of \ac{TSE} from a mixture of speakers by informing it about the target speaker's \ac{DoA} via a low-cost extension of the network (cf.~Sec.~\ref{subsubsec:doa}). Furthermore, we explicitly exploit the provided multichannel information by replacing \ac{2D} with \ac{3D} convolutional layers at the beginning of the network (cf.~Sec.~\ref{subsubsec:3dconv}). We show that these extensions allow to identify the target speaker and enhance its signal in the presence of interfering speakers, rendering the proposed \ac{iCOSPA} a flexible spatial filter. 
For reverberant scenarios, several options for the target signal used for training the neural filter exist. We examine how the spatial filtering capability is affected by different target signals and also show how the evaluation metrics depend on the choice of reference, i.e., clean signal, used for their computation. 

We present the proposed method and discuss the training target signal in Sec.~\ref{sec:method}. In Sec.~\ref{subsec:setup}, we detail our experimental setup and present and discuss the corresponding results in Sec.~\ref{subsec:results}. Sec.~\ref{sec:conclusion} concludes the paper.
%
\section{Providing spatial information for COSPA}
\label{sec:method}
In the following, we briefly introduce the \ac{COSPA} framework and explain how it is modified to exploit spatial information for \ac{TSE} (Sec.~\ref{subsec:iCOSPA}). A discussion on the spatial selectivity obtained by appropriate target signals for multichannel processing follows in Sec.~\ref{subsec:method_targets}.
\subsection{Extension of COSPA}
\label{subsec:iCOSPA}
We consider a signal with $M$ microphone channels, captured by an arbitrary microphone array, where the signal at microphone $m$ in time-frequency bin $(\tau, f)$ is given by
\vspace{-6pt}
\begin{equation}
    X_m(\tau, f) = D_m(\tau, f) +\sum_{i=1}^{I}U_{im}(\tau, f) + N_m(\tau, f).
\end{equation}

\vspace{-6pt} %
\noindent $D_m(\tau, f)=H^*_m(\tau, f)S(\tau, f)$ denotes the desired speaker's signal at microphone $m$ based on the acoustic transfer function $H_m(\tau, f)$ from the desired speaker to the $m$-th microphone. $U_{im}$ denotes the contribution of interfering speaker $i$, $i=1, \dots, I$, at microphone~$m$ and $N_m$ is additional low-level sensor noise. The goal is to suppress the interfering speakers and to extract the source signal $S$, or a reverberant image of it, with minimal distortions. Thus, dereverberation is not explicitly addressed in this paper. We allow different target speakers and speaker positions across utterances but assume that the speakers' identities and positions remain static within one utterance. 

In \cite{Halimeh2022}, \ac{COSPA} was introduced for multichannel denoising. This framework consists of an \textit{encoder}, which, using a subnetwork denoted by \crunet, estimates a single-channel mask that is applied to all input channels and subsequently includes feature compression, a \textit{compandor} which effectuates multichannel processing, i.e., allows to process each channel differently, and a \textit{decoder} which outputs an individual mask $\mathcal{M}_m$ for each channel. In this paper, we modify \ac{COSPA} to the problem of \ac{TSE} by adding \ac{DoA} information.
\begin{figure*}[t] 
	\centering
    \input{2_Figure_architecture.pdf_tex}
    \caption{Architecture of iCOSPA (adapted from \cite{Halimeh2022}). Differences to COSPA are marked in red.}
    \label{fig:architecture}
\end{figure*}
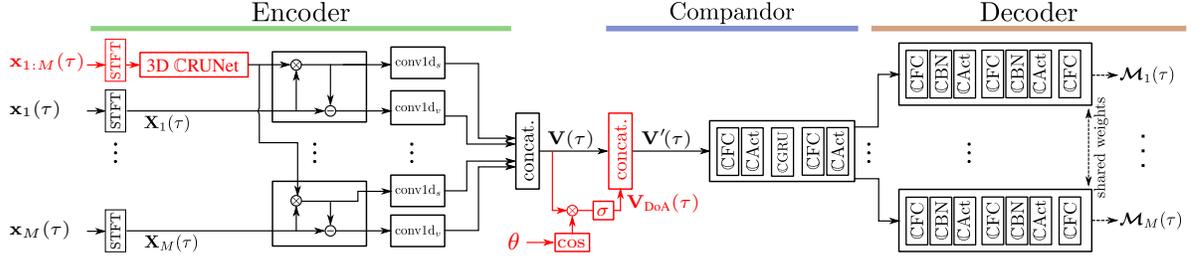
\subsubsection{Incorporating DoA information}
\label{subsubsec:doa}
For \ac{TSE} from multispeaker mixtures, the network needs to be informed about which speaker is the target speaker. This can be achieved by feeding characterizing (instantaneous) information about the target speaker into the network for guidance. This can either be positional, i.e., \ac{DoA}, information \cite{Chen2018, Gu2019}, or identity information such as an adaptation utterance spoken by the target speaker \cite{Zmolikova2019, Han2021, Elminshawi2022, Hsu2022}. %
The feature extraction involved in using an adaptation utterance for guidance is usually realized by an auxiliary neural network (e.g., \cite{Zmolikova2019}), which significantly increases the size of the overall neural network, while features based on \ac{DoA} information can often be extracted without a neural network (e.g., \cite{Chen2018}). In this paper, we want to focus on the spatial filtering aspect of \ac{COSPA}, for which the \ac{DoA} information is expected to be more relevant than the speaker identity. Considering these aspects and methods from literature, we find that exploiting \ac{DoA} information and adding it to the network using scaled activations \cite{Zmolikova2019} works well for \ac{COSPA} and also keeps the computational overhead small.
Therefore, as depicted in Fig.~\ref{fig:architecture}, for the proposed \textit{\ac{iCOSPA}} we scale the encoder output $\mathbf{V}$ by the cosine of the \ac{DoA} $\theta$ relative to the microphone array axis, pass it through a sigmoid activation function $\sigma$ and append it as a guiding signal $\mathbf{V}_{\text{DoA}}$ to the original vector $\mathbf{V}$ to obtain the input to the compandor $\mathbf{V'} = [\mathbf{V}^\top \mathbf{V}_{\text{DoA}}^\top]^\top$.
This differs from \cite{Zmolikova2019} as the original features are preserved. 
We add the guiding signal path in front of the compandor as this is the only part of \ac{COSPA} that can process the different channels differently and where the \ac{DoA} information can be beneficially used to increase the spatial selectivity.
Given the many proven methods for \ac{DoA} estimation in literature \cite{DiBiase2001, Pertila2018}, we assume that the utterance-wise \ac{DoA} of the target speech is available with sufficient precision. The proposed approach doubles the size of the input of the first compandor layer but does not add other trainable parameters.
\vspace{-12pt}
\subsubsection{3\D\ convolutional layer for multichannel processing}
\label{subsubsec:3dconv}
In order to leverage multichannel information in the encoder, we use all channels to compute the single-channel mask in the encoder. We replace the complex-valued \ac{2D} convolutions in the \crunet\ of \ac{COSPA} \cite{Halimeh2022}, which operate along the time and the frequency axis of the input signal, by complex-valued \ac{3D} convolutions, which additionally extract features across all input channels. \ac{3D} convolutional layers have been used in biomedical and video signal processing \cite{Ji2010, Ronneberger2016}, but not in audio signal processing. In the case of multichannel audio signal processing, extracting features across all three dimensions of the input tensor is intuitively beneficial as information about the spatial setup of the acoustic scene is encoded in the phase differences of the signals captured by the different channels, which also calls for complex-valued networks when operating in the time-frequency domain. Due to the relatively small kernel sizes of convolutional layers, replacing the \ac{2D} convolutions by \ac{3D} convolutions does not increase the size of the network significantly. 
\subsection{Training targets for spectro-spatial filtering}
\label{subsec:method_targets}
For joint spectro-spatial filtering with neural networks, several training targets, notably the clean source signal at the first microphone \cite{Li2022}, the time-aligned dry clean source signal \cite{Tesch2022}, and a \ac{MVDR BF}-filtered version of the source signal at the microphones \cite{Halimeh2022} have been used in the literature. %
Further potentially advantageous target signals can be obtained by processing the clean reverberated signal with other beamformers, e.g., a simple \ac{DSB} to align the time of arrival of the microphone channels, or by modifying the source image at the microphone otherwise (e.g., removing the late reverberation). 
To optimally approximate any of these targets, the network has to reflect different spatial and spectral filters. For the dry source signal as target, dereverberation has to be achieved as part of the spatial filtering, while the beamformer-filtered signals as targets tolerate certain amounts of reverberation. To estimate the source image at the first microphone, the network has to implicitly estimate relative transfer functions. %
Therefore, the choice of the training target is expected to have a significant impact on the spectro-spatial filtering patterns learned by the network.

In the following, we experimentally investigate the spatial filtering behavior of \ac{iCOSPA} when guided by the additional \ac{DoA} information. Furthermore, it is investigated how the training target influences the spatial filtering characteristics of \ac{iCOSPA}. As target signals we consider: the \ac{MVDR BF}-filtered reverberant source signal (\textit{mvdr}), the \ac{DSB}-filtered reverberant source signal (\textit{dsb}), the reverberant source signal as captured by the first microphone adding a small delay to avoid a theoretically possible need for noncausal processing (\textit{mic1}), and the dry source signal (\textit{dry}), time-aligned with the signal at the first microphone.
\vspace{-4pt}
\section{Experimental validation}
\label{sec:experiments}
In the experimental evaluation, we compare the proposed informed \ac{iCOSPA} with the uninformed \ac{COSPA} for \ac{TSE}. \ac{COSPA} can only know which \ac{DoA} corresponds to the target speaker when trained accordingly. Hence, we train and test \ac{COSPA} for four exemplary distinct positions of the target speaker separately, while training \ac{iCOSPA} on a wide range of target \ac{DoA} simultaneously (cf.~Sec.~\ref{subsec:setup}). We then show in Sec.~\ref{subsubsec:PoC} that \ac{iCOSPA} performs very similarly to \ac{COSPA}, but is much more flexible in deployment as a single trained network can be used for a wide range of different \acp{DoA}. Furthermore, in Sec.~\ref{subsubsec:results_targets}, we discuss the spatial selectivity patterns learned by \ac{iCOSPA} based on various training target signals and show the relevance of the characteristics of the reference signal for the computation of the evaluation metrics in reverberant scenarios.%
\vspace{-5pt}
\subsection{Setup and evaluation metrics}
\label{subsec:setup}
In our experiments, we consider scenarios with one target and $I=2$ interfering speakers. For each scenario, the room dimensions (in the range of [$4\ldots8$, $4\ldots8$, $1\ldots4$]\,m), the reverberation time ($0.2\ldots0.5$\,s), and the position of the microphone array are sampled randomly. We use a uniform linear array with $M=3$ microphones and an inter-microphone distance of $4$\,cm. In all scenarios, the target speaker is positioned $0.3\ldots1.5$\,m from the array. The interferers are placed randomly in the room but at least $0.3$\,m away from the walls and with an angular distance of at least $10\degree$  to the target speaker to both sides. The \acp{DoA} of all speakers are confined to $0\degree\ldots180\degree$ between the two endfire positions ($0$\degree{} and $180$\degree) of the array. Each source signal is convolved with its corresponding room impulse response generated by the image method \cite{Habets2010} and the interferers are scaled such that the target-to-interferer power ratio ranges from $-5$ to $5$\,dB. White noise is added to the mixture signals at a signal-to-noise ratio of $30\ldots60$\,dB. The speech signals are taken from the TIMIT database \cite{timit} and sampled at $16$\,kHz to form sequences of 7\,s duration.
For training, we generate five training datasets. Four training datasets contain $3000$ sequences each, where the target speaker is positioned at a fixed \ac{DoA} $\theta \in [0, 30, 60, 90]\degree$, respectively. These datasets are used to train \ac{COSPA} for \ac{TSE}, where the information about the target speaker's position has to be fixed in order to be learnable. The fifth dataset contains $250$ samples per target \ac{DoA} covering the range from $0\degree$ to $180\degree$ in steps of $5\degree$ for training \ac{iCOSPA}. For testing, we create $250$ sequences per target \ac{DoA} $\theta \in [0, 30, 60, 90]\degree$ with speakers disjoint from the training dataset and otherwise keep the same settings as for training.

In the following, we use the names \ac{COSPA} and \ac{iCOSPA} to discuss aspects of general relevance to the networks and use the \ac{MVDR BF}-filtered source signal as target if not stated otherwise. We append \textit{-mvdr/dsb/mic1/dry} to the name when discussing the networks trained on specific targets. For the mic1-target experiments, the estimated mask for each channel is constrained to a maximum magnitude of $1/M$ to ensure that the network does not collapse into a single-channel method. We assume the 3D convolutional layer in the \crunet\ of \ac{iCOSPA} to be the default setting and append \textit{-2D} when discussing \ac{iCOSPA} with the 2D \crunet.
We parameterize both \ac{COSPA} and \ac{iCOSPA} as in \cite{Halimeh2022}, with the exception of the additional convolutional kernels of size \{2, 2, 1, 1\} along the channel axis in the four modules of \ac{iCOSPA}'s \crunet. Furthermore, the input of \ac{iCOSPA}'s compandor, $\mathbf{V'}$, is twice the size of that of \ac{COSPA}, $\mathbf{V}$, due to the appended \ac{DoA} information $\mathbf{V}_{\text{DoA}}$. All network sizes are given in Table~\ref{Table:ResultsDoA}. We use signal frames of length $1024$ with a shift of $512$ samples. 
In Sec.~\ref{subsec:results}, we show beampatterns for \ac{iCOSPA} that were generated as described in \cite{Halimeh2022}.

For comparison, we provide results from the \ac{EaBNet} introduced in \cite{Li2022} as an alternative neural spectro-spatial filter. We keep the settings of the \ac{EaBNet}, including the frame length of 320 samples and the mic1 training target, as published without using an extra postfilter and train and test the network on the same datasets as \ac{COSPA}.

The performance of the methods is measured by \ac{PESQ} \cite{pesq}, \ac{ESTOI} \cite{Jensen2016} and the \ac{SIR} \cite{Vincent2006}. We provide the performance metrics as the difference between metrics of the estimated signal and metrics of the mixture signal at the first microphone averaged over the respective test dataset(s).
The discussion of which target to use for training, also raises the question which reference, i.e., clean, signal to use to compute the objective evaluation metrics. 
Here, we present the evaluation metrics based on both the dry source signal (\textit{dry}) and the reverberated source image at the first microphone (\textit{mic1}) as reference to illustrate how the presence or absence of reverberation in the reference signal affects the performance metrics for different target signals. Both signals are time-aligned with the estimate.
\vspace{-6pt}
\subsection{Results and Discussion}
\label{subsec:results}
We split the presentation of the results into two parts: In Sec.~\ref{subsubsec:PoC}, we show the \ac{TSE} performance of \ac{iCOSPA} compared to \ac{COSPA} and the \ac{EaBNet}. In Sec.~\ref{subsubsec:results_targets}, we discuss the influence of the training targets introduced in Sec.~\ref{subsec:method_targets} on the spatial selectivity of \ac{iCOSPA}. 
\vspace{-6pt}
\subsubsection{Performance evaluation}
\label{subsubsec:PoC}
\input{4_Table_DoA}
In Table~\ref{Table:ResultsDoA}, we present $\Delta$\ac{PESQ}, $\Delta$\ac{ESTOI} and $\Delta$\ac{SIR} for \ac{COSPA}, \ac{iCOSPA}, \ac{iCOSPA}-2D and the \ac{EaBNet} for four different target \acp{DoA}. As noted in Sec.~\ref{subsec:setup}, performance metrics are computed based on a dry and a reverberated signal. It can be seen that \ac{iCOSPA}, even though trained for a wide range of target \acp{DoA}, for all metrics performs very similarly as \ac{COSPA} which was trained for each test-\ac{DoA} specifically. This confirms that \ac{iCOSPA} beneficially uses the provided \ac{DoA} information for finding the correct target speaker. Hence, with only a very slight increase in model size, \ac{iCOSPA} is able to flexibly extract a target speaker from any direction.

Comparing \ac{iCOSPA} and \ac{iCOSPA}-2D, it can be seen that the two networks perform similarly, with \ac{iCOSPA} outperforming \ac{iCOSPA}-2D in some cases. The differences in performance are too small to make definite assertions. In our experiments, we noticed that the 3D convolutional layer seems to provide the most benefit in low SNR scenarios with high reverberation times. Further investigation on the contribution of the 3D convolutional layer is left for future work. We also investigated the robustness of \ac{iCOSPA} to \ac{DoA}-estimation errors, given that in a real scenario the exact \ac{DoA} might not be available. Adding uniformly sampled estimation noise from $-10\degree$ to $10\degree$ to the true \ac{DoA} in testing does not notably influence the performance of \ac{iCOSPA}, given that no other speakers are positioned in this region in our experiments.

It can be noted that the metrics computed on the two reference signal versions differ notably, which shows that the choice of the reference signal strongly influences the results. Since the metrics are computed based on the signal at the first microphone, which is closest to a \ac{DoA} of $0$\degree{} in our setup, the performance metrics can vary across \acp{DoA} for all methods. According to informal listening tests, \ac{COSPA} and \ac{iCOSPA} produce very similar results for all \acp{DoA}\footnote{Examples are provided at \texttt{https://github.com/LMSAudio/}.}. %

Furthermore, Table~\ref{Table:ResultsDoA} shows significantly different performance for the \ac{EaBNet} compared to \ac{COSPA}. While metrics based on the dry reference signal are notably lower, metrics based on the reverberated reference signal are significantly higher than those for \ac{COSPA}. This is most likely due to the different training targets of the two networks (mic1 for the \ac{EaBNet} and mvdr for (i)\ac{COSPA}). We discuss this effect of different training targets on the spatial selectivity and the evaluation metrics for \ac{iCOSPA} in Sec.~\ref{subsubsec:results_targets}.
\vspace{-8pt}
\subsubsection{Influence of training target on spatial filtering}
\label{subsubsec:results_targets}
\input{4_Figure_targets.tex}%
\vspace{-2pt} 
Building on the discussion of training targets for spectro-spatial filtering in Sec.~\ref{subsec:method_targets}, Fig.~\ref{fig:targets} shows exemplary beampatterns for \ac{iCOSPA}. It can be seen that the beampattern for the network trained on the \ac{MVDR BF}-filtered signal shows the most pronounced and precise beampattern. The \ac{DSB}-filtered target also leads to a distinct beampattern, which indicates that the network is able to learn the spatial selectivity patterns imposed on the target signals in the case of the two beamformers. As can be seen in Fig.~\ref{subfig:mic1}, using the desired signal at the first microphone as training target also leads to spatial awareness but with less suppression in the non-target directions, while the beampattern of the network trained on the dry source signal shown in Fig.~\ref{subfig:dry} reflects the spatial awareness of the two beamformer-filtered signals with stronger overall suppression. Moreover, Fig.~\ref{subfig:dry} shows that spatial selectivity can also be achieved by \ac{iCOSPA} without enforcing a specific spatial filtering process by the target signal. The attenuation in Fig.~\ref{subfig:dry} is due to the more aggressive filtering required to suppress the reverberation to attain the dry source target. 
The less strong suppression for the mic1-target in Fig.~\ref{subfig:mic1} may be attributed to the reverberation contained in the target, which corresponds to signal parts from all directions. In conclusion, the training target not only provides the network with the ideal characteristics of the estimated signal, but also influences the spatial filter represented by \ac{iCOSPA}. In any case, however, \ac{iCOSPA} identifies a beampattern pointing to the correct source \ac{DoA}.

\input{4_Table_average}
Table~\ref{Table:Results} summarizes the performance of \ac{iCOSPA} trained on various target signals. It can be seen that the evaluation metrics vary strongly with the training target and the reference signal. For the metrics based on the dry source signal as reference, the network trained on the first microphone channel performs worst, while the \ac{MVDR BF}-filtered target and the dry source target compete for the best performance. For the metrics based on the reverberated source signal, the \ac{DSB}-filtered target and the mic1-target give the best results, possibly because, compared to the other targets, they retain more of the unprocessed reverberation and hence correspond better to the reverberated reference signal.

The listening impression matches the interpretation of the beampatterns and the results presented in Table~\ref{Table:Results}. \ac{iCOSPA}-mvdr and \ac{iCOSPA}-dsb generate similar results, while the results from \ac{iCOSPA}-mic1 preserve more reverberation and with that also more of the interfering signals. The results generated from \ac{iCOSPA}-dry show the best interferer suppression but also contain some artefacts. This correlates with the strong suppression visible in Fig.~\ref{subfig:dry}. In general, all \ac{iCOSPA} variants generate very good speech quality for the target speaker and differ mostly in the suppression of the interferers, which is also reflected in the beampatterns in Fig.~\ref{fig:targets}. The decision on the `best version' will still depend on the scenario, as e.g., for sources close to the first microphone dereverberation will not be as desirable as for distant sources in a highly reverberant room, for which training with clean sources may be preferable.

Finally, in experiments with the \ac{EaBNet}, some dependency of the spatial selectivity of the network on the target signal can also be observed.
This supports the findings that choosing the training target for spectro-spatial filtering impacts not only the performance, but also the interpretability of a method. The generalizability of the results discussed above to other neural network architectures will be addressed in further work.
\vspace{-2pt}
\section{Conclusion and outlook}
\label{sec:conclusion}
\vspace{-2pt} In this paper, we presented \ac{iCOSPA}, an informed extension of \ac{COSPA} for \ac{TSE} that exploits additional \ac{DoA} information. Moreover, we adapted the 2D \crunet\ in \ac{iCOSPA}'s encoder to use a 3D convolutional layer. We showed that \ac{iCOSPA} uses the provided \ac{DoA} information to form a flexible spatial filter which reliably extracts the target speaker. 
The main contributions of this paper are the analysis and interpretation of the influence of the training target on the spatial filtering behavior of \ac{iCOSPA}, and the demonstration of the impact of the reference signal on the evaluation metrics. We found that the \ac{iCOSPA} architecture allows to learn beampatterns directed towards the target speaker even if only a dry source signal is used as training target. When using a target signal that results from spatial filtering, the spatial selectivity of the resulting beamformers is significantly more pronounced.

A more comprehensive evaluation of the 3D convolutional layer for multichannel audio signal processing is planned for future work. This includes analyzing the effect the 3D processing has on mask estimation in \ac{iCOSPA}'s encoder, and investigating the impact of equalizing the phase differences between microphones before using the 3D convolutional layer.
%

\bibliographystyle{IEEEbib}
\bibliography{refs}

\end{document}

%% file: 2_Figure_architecture.pdf_tex
%

 	\begingroup%
	\providecommand\color[2][]{%
		\errmessage{(Inkscape) Color is used for the text in Inkscape, but the package 'color.sty' is not loaded}%
		\renewcommand\color[2][]{}%
	}%
    \setlength{\unitlength}{1cm}
    \setlength{\fboxsep}{0pt}
	
	\begin{picture}(15, 3.34)%
    	\put(0,0){\includegraphics[height=3.34cm]{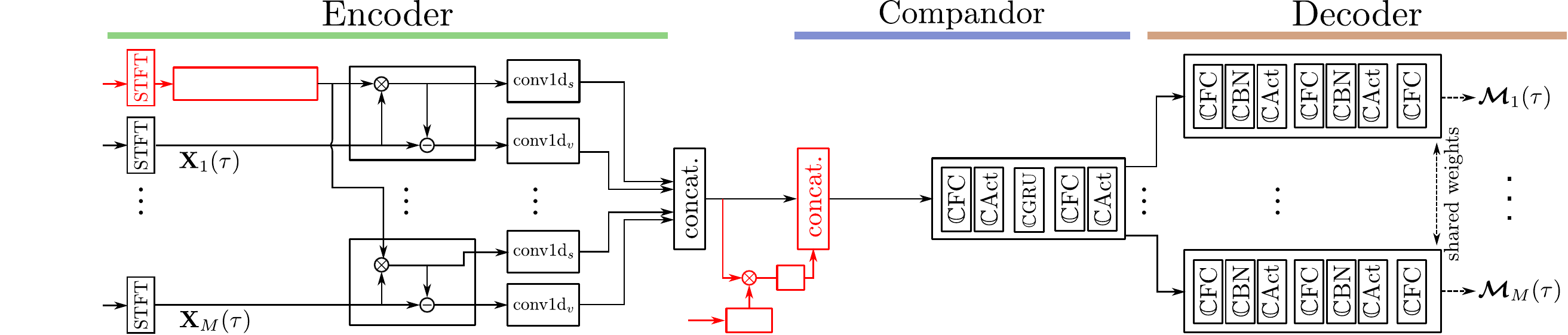}}%

    	\put(6.62,.05){\color[rgb]{1,0,0} \small $\theta$}%
        \put(7.8,.5){\color[rgb]{1,0,0} \scriptsize $\sigma$}%
        \put(8.2,.6){\color[rgb]{1,0,0} \scriptsize $\mathbf{V}_{\text{DoA}}(\tau)$}%
        \put(7.28,.07){\color[rgb]{1,0,0} \scriptsize $\cos$}%
        \put(0,2.48){\color[rgb]{1,0,0} \scriptsize $\mathbf{x}_{1:M}(\tau)$}%
        \put(1.82,2.42){\color[rgb]{1,0,0} \scriptsize 3D \crunet}%
        \put(0,1.85){\scriptsize $\mathbf{x}_{1}(\tau)$}%
        \put(0,0.24){\scriptsize $\mathbf{x}_{M}(\tau)$}%
        \put(7.15, 1.45){\scriptsize $\mathbf{V(\tau)}$}
        \put(8.4, 1.45){\scriptsize $\mathbf{V'\!(\tau)}$}
	
	\end{picture}%
	\endgroup%

%% file: 4_Table_DoA.tex
\newcommand{\STAB}[1]{\begin{tabular}{@{}c@{}}#1\end{tabular}}

\begin{table}[t]
\setlength{\tabcolsep}{3pt} 
\renewcommand*{\arraystretch}{1.1}
\centering
\caption{Performance metrics for \ac{TSE} for various target \acp{DoA}. Metrics are based on the \textit{dry/mic1} reference signals. Note that the results for \ac{iCOSPA}(-2D) are all obtained from the same network, while the results for \ac{COSPA} and the \ac{EaBNet} each come from four separately trained networks. The EaBNet is trained on the mic1 target, the other networks on the mvdr target.} 
\scalebox{0.82}{
\begin{tabular}{ccccccc}
\toprule
 & \multirow{2}{*}{Model} & \# Param. & \multirow{2}{*}{$0\degree$} & \multirow{2}{*}{$30\degree$} & \multirow{2}{*}{$60\degree$} & \multirow{2}{*}{$90\degree$} \\
 & & [million] & & & & \\
 \hline
\multirow{4}{*}{\STAB{\rotatebox[origin=c]{90}{$\Delta$PESQ}}} & EaBNet & $2.8$ & $0.18/0.45$ & $0.21/0.50$  & $0.16/0.39$  & $0.14/0.37$ \\ 
& COSPA & $2.1$ & $0.26/0.24$  & $0.25/0.28$  & $0.16/0.23$  & $0.15/0.29$ \\
& iCOSPA-2D & $2.5$ & $0.23/0.19$ &$0.23/0.25$  & $0.15/0.20$ & $0.13/0.23$ \\
& iCOSPA & $2.6$ & $0.24/0.19$ &$0.24/0.25$  & $0.17/0.23$ & $0.14/0.26$ \\
\hline 
\multirow{4}{*}{\STAB{\rotatebox[origin=c]{90}{$\Delta$ESTOI}}} & EaBNet & $2.8$ & $0.13/0.13$& $0.13/0.13$ & $0.10/0.11$ & $0.10/0.12$ \\ 
 & COSPA & $2.1$ & $0.19/0.04$ & $0.17/0.04$ & $0.11/0.05$ & $0.11/0.08$ \\
& iCOSPA-2D & $2.5$ & $0.18/0.03$ & $0.16/0.05$ & $0.10/0.04$ & $0.09/0.07$ \\
& iCOSPA & $2.6$ & $0.19/0.03$ & $0.17/0.05$ & $0.11/0.05$ & $0.09/0.07$ \\
\hline
\multirow{4}{*}{\STAB{\rotatebox[origin=c]{90}{$\Delta$SIR [dB]}}} & EaBNet & $2.8$ & $8.88/12.03$& $8.78/12.37$ & $9.00/11.59$ & $8.63/12.04$ \\ 
 & COSPA & $2.1$ & $11.74/8.99$ & $11.05/9.23$ & $9.98/9.50$ & $8.79/9.75$ \\
& iCOSPA-2D & $2.5$ & $11.43/8.63$ & $10.86/9.04$ & $9.67/9.22$ & $8.71/9.84$ \\
& iCOSPA & $2.6$ & $11.66/8.88$ & $10.93/9.10$ & $9.82/9.34$ & $8.66/9.68$ \\
\bottomrule
\end{tabular}
}
\label{Table:ResultsDoA}
\end{table}

%% file: 4_Figure_targets.tex
        
%
%
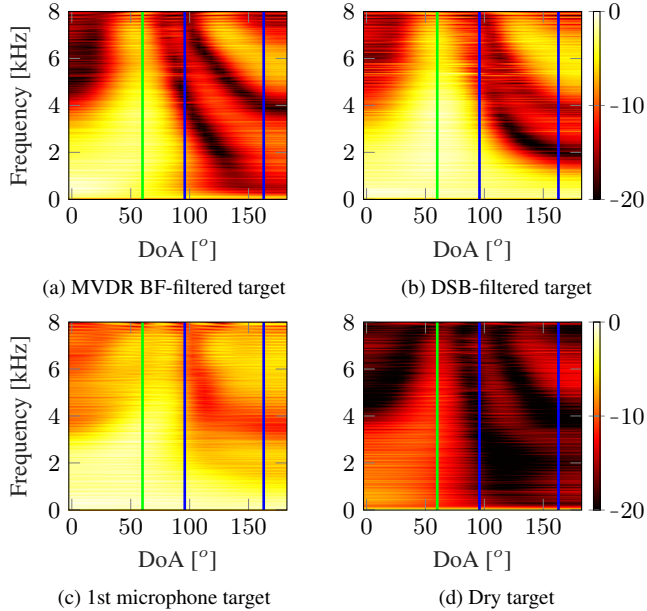
\begin{figure}[t]
    \input{mvdr_DOA60_sample14.tikz}
    \input{dsb_DOA60_sample14.tikz}
        	 
    \input{mic1_DOA60_sample14.tikz}
    \input{dry_DOA60_sample14.tikz}  	
        	
    \caption{Exemplary beampatterns for iCOSPA trained on different target signals. green line: target speaker; blue lines: interferers}
    \label{fig:targets} 	
\end{figure}

%% file: mvdr_DOA60_sample14.tikz
%
%
\begin{subfigure}[t]{0.5\columnwidth}
\begin{tikzpicture}

\begin{axis}[%
width=2.9cm, 
ylabel near ticks,
xlabel near ticks,
restrict x to domain*=-2.57:183,
restrict y to domain*=-0.01:8,
scale only axis,
point meta min=-20,
point meta max=0,
axis on top,
xmin=-2.57142857142857,
xmax=182.571428571429,
xlabel style={font=\color{white!15!black}},
xlabel={DoA [$^{o}$]},
ymin=-0.00780487060546875,
ymax=7.99999237060547,
ylabel style={font=\color{white!15!black}},
ylabel={Frequency [kHz]},
axis background/.style={fill=white},
colormap/hot2,
]
\addplot [forget plot] graphics [xmin=-2.57142857142857, xmax=182.571428571429, ymin=-0.00780487060546875, ymax=7.99999237060547] {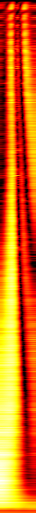};
\addplot [color=green, line width=1.0pt, forget plot]
  table[row sep=crcr]{%
60	0\\
60	8000\\
};
\addplot [color=blue, line width=1.0pt, forget plot]
  table[row sep=crcr]{%
95.8440778319705	0\\
95.8440778319705	8000\\
};
\addplot [color=blue, line width=1.0pt, forget plot]
  table[row sep=crcr]{%
162.864215677866	0\\
162.864215677866	8000\\
};
\end{axis}
\end{tikzpicture}%
\vspace{-3pt}
\subcaption{MVDR BF-filtered target}
\label{subfig:mvdr}
\end{subfigure}

%% file: dsb_DOA60_sample14.tikz
%
%
\begin{subfigure}[t]{0.5\columnwidth}
\begin{tikzpicture}

\begin{axis}[%
width=2.9cm, 
ylabel near ticks,
xlabel near ticks,
restrict x to domain*=-2.57:183,
restrict y to domain*=-0.01:8,
scale only axis,
point meta min=-20,
point meta max=0,
axis on top,
xmin=-2.57142857142857,
xmax=182.571428571429,
xlabel style={font=\color{white!15!black}},
xlabel={DoA [$^{o}$]},
ymin=-0.00780487060546875,
ymax=7.99999237060547,
ylabel style={font=\color{white!15!black}},
axis background/.style={fill=white},
colormap/hot2,
colorbar,
colorbar style={          
            width=0.03*\pgfkeysvalueof{/pgfplots/parent axis width},
            ytick={-20, -10, 0},
            yticklabels={$\shortminus20$, $\shortminus10$, $0$},
            xshift=-0.15cm,
        },
colorbar/width=2.5mm,
]
]
\addplot [forget plot] graphics [xmin=-2.57142857142857, xmax=182.571428571429, ymin=-0.00780487060546875, ymax=7.99999237060547] {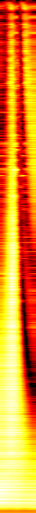};
\addplot [color=green, line width=1.0pt, forget plot]
  table[row sep=crcr]{%
60	0\\
60	8000\\
};
\addplot [color=blue, line width=1.0pt, forget plot]
  table[row sep=crcr]{%
95.8440778319705	0\\
95.8440778319705	8000\\
};
\addplot [color=blue, line width=1.0pt, forget plot]
  table[row sep=crcr]{%
162.864215677866	0\\
162.864215677866	8000\\
};
\end{axis}
\end{tikzpicture}%
\vspace{-3pt}
\subcaption{DSB-filtered target}
\label{subfig:dsb}
\end{subfigure}

%% file: mic1_DOA60_sample14.tikz
%
%
\begin{subfigure}[t]{0.5\columnwidth}
\begin{tikzpicture}

\begin{axis}[%
width=2.9cm, 
ylabel near ticks,
xlabel near ticks,
restrict x to domain*=-2.57:183,
restrict y to domain*=-0.01:8,
scale only axis,
point meta min=-20,
point meta max=0,
axis on top,
xmin=-2.57142857142857,
xmax=182.571428571429,
xlabel style={font=\color{white!15!black}},
xlabel={DoA [$^{o}$]},
ymin=-0.00780487060546875,
ymax=7.99999237060547,
ylabel style={font=\color{white!15!black}},
ylabel={Frequency [kHz]},
axis background/.style={fill=white},
colormap/hot2,
]
\addplot [forget plot] graphics [xmin=-2.57142857142857, xmax=182.571428571429, ymin=-0.00780487060546875, ymax=7.99999237060547] {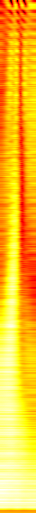};
\addplot [color=green, line width=1.0pt, forget plot]
  table[row sep=crcr]{%
60	0\\
60	8000\\
};
\addplot [color=blue, line width=1.0pt, forget plot]
  table[row sep=crcr]{%
95.8440778319705	0\\
95.8440778319705	8000\\
};
\addplot [color=blue, line width=1.0pt, forget plot]
  table[row sep=crcr]{%
162.864215677866	0\\
162.864215677866	8000\\
};
\end{axis}
\end{tikzpicture}%
\vspace{-3pt}
\subcaption{1st microphone target}
\label{subfig:mic1}
\end{subfigure}

%% file: dry_DOA60_sample14.tikz
%
%
\begin{subfigure}[t]{0.5\columnwidth}
\begin{tikzpicture}

\begin{axis}[%
width=2.9cm, 
ylabel near ticks,
xlabel near ticks,
restrict x to domain*=-2.57:183,
restrict y to domain*=-0.01:8,
scale only axis,
point meta min=-20,
point meta max=0,
axis on top,
xmin=-2.57142857142857,
xmax=182.571428571429,
xlabel style={font=\color{white!15!black}},
xlabel={DoA [$^{o}$]},
ymin=-0.00780487060546875,
ymax=7.99999237060547,
ylabel style={font=\color{white!15!black}},
axis background/.style={fill=white},
colormap/hot2,
colorbar,
colorbar style={          
            width=0.03*\pgfkeysvalueof{/pgfplots/parent axis width},
            ytick={-20, -10, 0},
            yticklabels={$\shortminus20$, $\shortminus10$, $0$},
            xshift=-0.15cm,
        },
colorbar/width=2.5mm,
]
]
\addplot [forget plot] graphics [xmin=-2.57142857142857, xmax=182.571428571429, ymin=-0.00780487060546875, ymax=7.99999237060547] {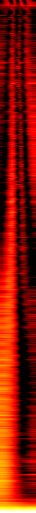};
\addplot [color=green, line width=1.0pt, forget plot]
  table[row sep=crcr]{%
60	0\\
60	8000\\
};
\addplot [color=blue, line width=1.0pt, forget plot]
  table[row sep=crcr]{%
95.8440778319705	0\\
95.8440778319705	8000\\
};
\addplot [color=blue, line width=1.0pt, forget plot]
  table[row sep=crcr]{%
162.864215677866	0\\
162.864215677866	8000\\
};
\end{axis}
\end{tikzpicture}%
\vspace{-3pt}
\subcaption{Dry target}
\label{subfig:dry}
\end{subfigure}

%% file: 4_Table_average.tex
\begin{table}[t]
\setlength{\tabcolsep}{6pt} 
\renewcommand*{\arraystretch}{1.1}
\centering
\caption{Performance metrics for \ac{TSE} averaged over test datasets. 
 Metrics are based on the \textit{dry/mic1} reference signals. The best value per metric is printed in bold.} 
\scalebox{0.84}{
\begin{tabular}{lccc}
\toprule
\multirow{2}{*}{Model}  & $\Delta$\ac{PESQ} & $\Delta$\ac{ESTOI} & $\Delta$\ac{SIR} \\
  & & & [dB]\\
\hline
iCOSPA-mvdr & $\mathbf{0.20}/0.23$ & $\mathbf{0.14}/0.05$ & $10.27/9.25$  \\
iCOSPA-dsb & $0.18/\mathbf{0.31}$  & $0.11/\mathbf{0.08}$ & $9.02/10.03$\\
iCOSPA-mic1 & $0.16/\mathbf{0.31}$ & $0.09/\mathbf{0.08}$ & $8.55/\mathbf{10.12}$ \\
iCOSPA-dry &  $0.19/0.09$ & $0.12/\shortminus0.01$ & $\mathbf{10.82}/8.42$\\
\bottomrule
\end{tabular}
}
\label{Table:Results}
\end{table}